\documentclass[sigconf,screen]{acmart}
\usepackage{enumitem}

\usepackage{tabularx}
\usepackage{geometry} 
\usepackage{booktabs} 
\usepackage{multirow}
\usepackage{pifont}
\usepackage{tabularx}
\usepackage{xcolor,colortbl}
\usepackage{wrapfig}
\usepackage{xspace}
\usepackage{tikz}
\usepackage{subcaption}
\usepackage{flushend}

\newcommand{\CI}{\texttt{g.CO$_2$/kWh}\xspace}
\newcommand{\C}{\texttt{g.CO$_2$}\xspace}
\newcommand{\SCI}{\texttt{SCI}\xspace}
\newcommand{\TSCI}{\texttt{tSCI}\xspace}
\newcommand{\OSCI}{\texttt{oSCI}\xspace}

\newcommand{\SA}{$\text{S}_\text{A}$\xspace}
\newcommand{\SB}{$\text{S}_\text{B}$\xspace}

\usepackage[many,theorems,skins,breakable]{tcolorbox}
\newtcolorbox{visionbox}[2][]{%
    colback=teal!10,
    coltitle=black,
    colframe=teal!50,
    fonttitle=\bfseries,
    title=#2, 
    sharp corners,
    rounded corners=southeast,
    boxrule=0pt,
    enhanced,
    drop fuzzy shadow,
    #1, 
    top=3pt,bottom=2pt,left=3pt,right=3pt
    }

\AtBeginDocument{%
  }

\copyrightyear{2024}
\acmYear{2024}
\setcopyright{rightsretained}
\acmConference[SoCC '24]{ACM Symposium on Cloud Computing}{November 20--22, 2024}{Redmond, WA, USA}
\acmBooktitle{ACM Symposium on Cloud Computing (SoCC '24), November 20--22, 2024, Redmond, WA, USA}
\acmDOI{10.1145/3698038.3698542}
\acmISBN{979-8-4007-1286-9/24/11}

\begin{document}
\sloppy

\title[Sunk Carbon Fallacy]{The Sunk Carbon Fallacy: Rethinking Carbon Footprint Metrics for Effective Carbon-Aware Scheduling}

\author{Noman Bashir}
\affiliation{%
  \institution{Massachusetts Institute of Technology}
  \city{}
  \country{}}

\author{Varun Gohil}
\affiliation{%
  \institution{Massachusetts Institute of Technology}
  \city{}
  \country{}}

\author{Anagha Belavadi}
\affiliation{%
  \institution{Massachusetts Institute of Technology}
  \city{}
  \country{}}

\author{Mohammad Shahrad}
\affiliation{%
  \institution{University of British Columbia}
  \city{}
  \country{}}

\author{David Irwin}
\affiliation{%
  \institution{University of Massachusetts Amherst}
  \city{}
  \country{}}

\author{Elsa Olivetti}
\affiliation{%
  \institution{Massachusetts Institute of Technology}
  \city{}
  \country{}}

\author{Christina Delimitrou}
\affiliation{%
  \institution{Massachusetts Institute of Technology}
  \city{}
  \country{}}

\renewcommand{\shortauthors}{Bashir et al.}

\begin{abstract}
The rapid increase in computing demand and its corresponding energy consumption have focused attention on  computing's impact on the climate and sustainability. Prior work proposes metrics that quantify computing's carbon footprint across several lifecycle phases, including its supply chain, operation, and end-of-life. Industry uses these metrics to optimize the carbon footprint of manufacturing hardware and running computing applications. 
Unfortunately, prior work on optimizing datacenters' carbon footprint often succumbs to the \emph{sunk cost fallacy} by considering embodied carbon emissions (a sunk cost) when making operational decisions (i.e., job scheduling and placement), which leads to operational decisions that do not always reduce the total carbon footprint. 

In this paper, we evaluate carbon-aware job scheduling and placement on a given set of servers for a number of carbon accounting metrics. Our analysis reveals state-of-the-art carbon accounting metrics that include embodied carbon emissions when making operational decisions can actually increase the total carbon footprint of executing a set of jobs. We study the factors that affect the added carbon cost of such suboptimal decision-making.  We then use a real-world case study from a datacenter to demonstrate how the sunk carbon fallacy manifests itself in practice.  Finally, we discuss the implications of our findings in better guiding effective carbon-aware scheduling in on-premise and cloud datacenters. 
\end{abstract}

\ccsdesc[500]{Hardware~Impact on the environment}
\ccsdesc[500]{Hardware~Emerging tools and methodologies}
\ccsdesc[500]{General and reference~Metrics}
\ccsdesc[500]{General and reference~Measurement}
\ccsdesc[500]{General and reference~Evaluation}
\ccsdesc[500]{General and reference~Empirical studies}

\keywords{Sustainable computing, operational carbon emissions, embodied carbon emissions, carbon footprint, sustainability, metrics, software carbon intensity, lifecycle carbon footprint, datacenters, server lifetime, scheduling, job placement, performance.}

\maketitle

\section{Introduction}
\label{sec:intro}
Computing's demand has experienced a meteoric rise over the last few decades with no signs of slowing down~\cite{Denning:2017:Growth}. Indeed, computing's demand is likely accelerating due to the recent emergence of generative artificial intelligence (AI) tools, such as  ChatGPT~\cite{Brown:2020:ChatGPT} and GitHub Copilot~\cite{Copilot2022GitHub}, that are computationally-intensive.  New AI tools promise to unlock a broad spectrum of innovative and useful applications. However, as marginal improvements in computing's energy efficiency shrink due to the deceleration of process scaling~\cite{Koomey:2015:Revised,Vahdat:2024:AsplosKeynote}, the ever-growing demand for computing power is poised to drive a proportional increase in its energy consumption. Computing's growing energy footprint has raised significant climate and sustainability concerns. Fortunately, the importance of improving computing's sustainability is gaining awareness~\cite{Bashir:2024:Climate,Strubell:2019:NLP,Wu:2022:SustAI}, with coordinated efforts from both industry and academia aimed at mitigating its climate impact~\cite{Radovanovic:2021:CarbonAwareCF, Wang:2023:Energy, Wu:2022:SustAI, Wang:2024:Designing, Bashir:2021:Enabling}.

Recent efforts to improve computing's sustainability have focused on quantifying and optimizing its carbon footprint across all stages of the computing lifecycle, from chip design and manufacturing~\cite{Acun:2022:Explorer, Gupta:2021:Chasing} to the operation of computer systems~\cite{Souza:2023:Ecovisor, Hanafy:2024:CarbonScaler, Li:2023:Clover} and the management of e-waste at the end of life~\cite{Switzer:2023:Junkyard}. 
The Greenhouse Gas (GHG) Protocol~\cite{GHG:1998:Protocol} distinguishes between different types of emissions: Scope 2 covers the GHG emissions related to electricity consumption in datacenters (often referred to as operational emissions), while Scope 3 includes emissions arising from chip manufacturing, the supply chain, and e-waste management (often referred to as embodied emissions). Prior work on quantifying computing's carbon footprint employs various metrics based on either operational emissions alone or a weighted combination of both operational and embodied emissions. 
A common approach in the literature is to aggregate the operational emissions from executing a job with a portion of the embodied emissions of the server running that job, where the server's embodied emissions are distributed across jobs based on the time and resources allocated to them. 
Notable examples include the Software Carbon Intensity (SCI), introduced by the Green Software Foundation~\cite{GSF:2021:SCI}, Computational Carbon Intensity~\cite{Switzer:2023:Junkyard}, and Sustainability Cost Rate~\cite{Gandhi:2023:Metrics}. While these metrics might use different terminology, the underlying principle remains consistent: a job's carbon footprint is the sum of its share of the hardware's embodied carbon emissions and the operational emissions generated during its execution.

In this paper, we focus on carbon-aware workload scheduling and job placement on datacenter servers. While embodied carbon metrics like \SCI are often proposed to guide operational decisions, such as scheduling and job placement, we argue that scheduling and procurement are orthogonal processes that operate on different timescales and should be optimized independently. Scheduling decisions determine which servers handle specific jobs and should target reducing the operational carbon footprint generated by actively running servers. In contrast, procurement decisions -- such as which servers to buy and when to replace them -- can affect the embodied carbon footprint associated with the hardware's manufacturing, which cannot be influenced when a job is being scheduled. These processes are separate: scheduling occurs continuously as jobs are assigned to servers, while procurement decisions are made at longer intervals based on hardware lifecycles.

Importantly, metrics like \SCI, which incorporate lifecycle emissions, typically account for only the emissions of the servers running the jobs, and ignore the embodied carbon of idle or unused servers. This oversight can lead to unintended consequences when optimizing for \SCI-like metrics in job scheduling by, paradoxically, increasing rather than decreasing a datacenter's overall carbon footprint, as it neglects the broader carbon impact of the entire server fleet, including unused hardware. Therefore, we demonstrate that optimizing \SCI-like metrics alone when scheduling may actually undermine the goal of minimizing a datacenter's total carbon footprint, highlighting the need for separate, independent optimization of scheduling and procurement.

The suboptimal outcomes of carbon-aware scheduling decisions based on \SCI-like metrics stem from a cognitive bias known as the \emph{sunk cost fallacy}. 
According to the \emph{principle of bygones}, a concept rooted in the principle of separability in standard economic theory, operators should base decisions solely on future possibilities, without being influenced by past expenditures or events that cannot be changed~\cite{Cubitt:2012:Bygones}. 
Applied to datacenter operations, this means that scheduling and job placement decisions should be made by focusing only on the current operational context by, in this case, ignoring the embodied carbon emissions that have already occurred. Since embodied emissions are fixed at procurement time, they cannot be altered by later operational choices. Thus, operators should aim to minimize operational carbon emissions from running jobs on the existing hardware rather than factoring in sunk embodied carbon.

Ignoring sunk costs is intuitive and supported by prior research~\cite{Gerstner:2014:NeuronalBook, Paschos:2019:Separation, Sun:2024:Timescale, Wei:2023:TimeDecoupling}. However, recent attempts to develop metrics that optimize computing's lifecycle carbon footprint inadvertently fall into the \emph{sunk carbon fallacy}, a variant of the sunk cost fallacy applied to carbon.  By incorporating embodied emissions into real-time scheduling decisions, these metrics conflate two separate processes: procurement and operation. Our illustrative example in~\autoref{sec:illustration} highlights how using \SCI as a scheduling metric can counterproductively increase a datacenter's overall carbon footprint, underscoring the importance of independently optimizing scheduling and procurement to optimize carbon efficiency.

The degree to which minimizing the total carbon footprint of a datacenter diverges from minimizing the sum of individual job-level lifecycle carbon using metrics like \SCI depends on several characteristics of datacenter infrastructure. 
One significant factor is the heterogeneity in servers’ performance relative to their operational and embodied carbon footprints. If all servers in the datacenter are homogeneous -- delivering similar performance for a given application -- then incorporating embodied carbon into a scheduling metric like \SCI would not affect the overall system-level carbon footprint. However, real-world datacenters often consist of heterogeneous servers, with differences arising from factors such as hardware age (new vs. old) and type (CPU vs. GPU).
For instance, older servers typically have a smaller embodied carbon footprint due to their earlier manufacture but often exhibit a higher operational carbon footprint than newer, more energy-efficient servers, as shown in~\autoref{fig:servers-example}. 
Additionally, while GPUs may excel in compute-intensive tasks, CPUs can sometimes deliver better performance per unit energy or carbon for specific workloads~\cite{Acun:2021:TrainEff}.

Our work focuses on heterogeneity in CPUs, which is significant enough to demonstrate that applying a one-size-fits-all metric like \SCI, which includes embodied carbon, can distort scheduling decisions in ways that increase the total carbon footprint. 
A second critical factor is datacenter utilization. If utilization is either very high or very low -- where either all servers are in use, or none are -- then the choice of scheduling metric will have little impact on the overall carbon footprint. However, at intermediate utilization levels, common in many datacenters, metrics like \SCI can lead to inefficient scheduling, increasing the total carbon footprint. In~\autoref{sec:illustration}, we further explore this discrepancy and evaluate the influence of these infrastructure factors on a datacenter's carbon footprint through concrete examples.

In illustrating how the \emph{sunk cost fallacy} manifests in carbon-aware scheduling, this paper makes the following contributions:

\vspace{0.1cm}
\noindent
\textbf{1 --} We demonstrate that metrics incorporating both embodied and operational carbon emissions, while seemingly comprehensive, can result in sub-optimal scheduling decisions. These metrics may paradoxically increase a datacenter’s total carbon footprint, contrary to their intended purpose. We also examine the key factors, such as datacenter utilization levels, operational carbon intensity, and embodied carbon amortization approaches that exacerbate these sub-optimal outcomes.

\vspace{0.1cm}
\noindent
\textbf{2 --} 
We evaluate three metrics, including those that prioritize operational carbon emissions or account for infrastructure-wide embodied carbon emissions more appropriately than \SCI. 
Through a real-world case study of an on-premise datacenter, we show that, under realistic workload assumptions, focusing on operational carbon emissions leads to more carbon-efficient scheduling outcomes, effectively reducing the total carbon footprint.

\vspace{0.1cm}
\noindent
\textbf{3 --} 
We provide practical guidelines for datacenter operators and users, detailing how to avoid the sunk carbon fallacy. Our recommendations include selecting metrics that accurately reflect the carbon costs relevant to operational decision-making, thereby optimizing for a lower total carbon footprint.

\section{Background and Motivation}
\label{sec:background}
This section provides an overview of efforts to improve computing's sustainability, the metrics used in carbon-aware optimizations, and the research gaps in understanding sustainability metrics.

\vspace{0.1cm}
\noindent
\textbf{Prior work on sustainable computing.}
There has been significant work  highlighting the environmental impact of computing~\cite{Strubell:2019:NLP} and understanding what it means for computing to be sustainable~\cite{Wu:2022:SustAI, Bashir:2023:HotAir}. 
Prior work has also analyzed various carbon accounting paradigms in the context of computing and highlighted the challenges in accounting for the carbon footprint of computing~\cite{Bashir:2023:Embodied, Dip:2024:Mirage}. 
Specifically, it examines how the values for embodied carbon~\cite{Bashir:2023:Embodied} and operational carbon intensity can be error-prone~\cite{Dip:2024:Mirage}. 
Recent work has also focused on quantifying operational and embodied carbon, as well as their tradeoff.  These efforts guide the architectural design process to reduce servers' overall lifecycle carbon footprint. 
There is also prior work on understanding the benefits and limitations of spatiotemporal workload scheduling for reducing carbon~\cite{sukprasert2023quantifying}.
An orthogonal body of work has focused on devising algorithms for carbon-aware workload shifting and building systems that enable the deployment of such carbon-aware algorithms~\cite{Souza:2023:Ecovisor, Hanafy:2024:CarbonScaler, Thiede:2023:CarbonContainers, Gsteiger:2024:Caribou}. 
Unfortunately, the real-world use of carbon-aware optimizations is limited, with a single example of carbon-aware workload shifting in hyperscalers~\cite{Radovanovic:2021:CarbonAwareCF}.

\vspace{0.1cm}
\noindent
\textbf{Metrics for sustainable computing.}
\label{sec:metrics}
Recent work has looked at various metrics that should be used to quantify and optimize computing's carbon footprint. 
\citet{Gandhi:2023:Metrics} propose sustainability metrics for datacenters, such as the amortized sustainability cost metric that attributes operational and embodied carbon to a job.
\citet{Switzer:2023:Junkyard} look at the end-of-life problem for computing hardware and propose a metric, called computational carbon intensity (CCI), for making component replacement and end-of-life decisions for computing hardware. 
The software industry has also focused on promoting and facilitating the development of green software, such as prior work done by the Green Software Foundation (GSF)~\cite{GSF:2021:SCI}.
GSF has proposed a metric, Software Carbon Intensity (SCI), that can quantify software's carbon footprint and enable software practitioners to make decisions that reduce it.

\vspace{0.1cm}
\noindent
\textbf{Limitations and research gaps.}
Prior work on carbon accounting has proposed various metrics to reduce computing's carbon footprint, which have spurred a debate on the usefulness and efficacy of the proposed metrics~\cite{Chien:2023:Metrics, Chien:2023:MetricsP2, Berger:2023:Metrics, Shenoy:2024:Optimizing}.
However, despite this being a critical problem, very little work has been done on analyzing the incentives each metric offers and the outcomes it yields. 
Recent work argues that devising a single metric that is simple, accurate, precise, and provides desirable incentives for optimizing decision-making across computing's entire lifecycle may not be possible~\cite{Taylor:2020:Metrics}. 
Simultaneously, it is challenging to evaluate all potential combinations. 
The total lifecycle carbon footprint includes the embodied carbon of all servers, operational carbon of idle servers, and carbon emissions of active servers running workloads. 
Procurement decisions and job scheduling choices matter, but they operate on different timescales: seconds to days for scheduling and months to years for procurement.  
This work focuses on carbon-aware workload scheduling in the public cloud and enterprise datacenters, aiming to reduce the carbon footprint added in this life stage.

\section{The Sunk Carbon Fallacy}
\label{sec:problem}
This section illustrates how the state-of-the-art carbon accounting metrics fall victim to the \emph{sunk carbon fallacy}, outlines factors that determine the impact of suboptimal decision-making on the total carbon footprint, and analyzes alternative metrics that yield better carbon-aware scheduling outcomes. 

\subsection{Analysis Setup}
\label{sec:setup}
The task of carbon-aware scheduling and job placement is to schedule a given set of jobs on a set of available servers to minimize the total carbon footprint of executing the jobs. 
In our illustrative example, we assume the following setup.

\vspace{0.05cm}
\noindent
\textbf{-- Scheduler.} The goal of the scheduler is to place a set of jobs onto a given set of servers to minimize the total carbon footprint of running all the jobs. 
The scheduler does not assume any information about the future arrival of jobs and their characteristics, and makes instantaneous decisions on job placement -- a setup commonly used by in-production schedulers, such as Borg~\cite{Bashir:2021:Limit, Tirmazi:2020:Borg}.

\vspace{0.05cm}
\noindent\textbf{-- Jobs.} The performance characteristics and energy consumption of jobs on the available servers are known. These characteristics can be obtained through profiling or increasingly common benchmark databases, such as MLPerf~\cite{Mattson:2020:MLPerf} and OpenBenchmarking Suite~\cite{OpenBenchmarking:2024:Suites}. 

\vspace{0.05cm}
\noindent\textbf{-- Servers.} The servers are not power proportional, i.e., they consume significant power at 0\% utilization~\cite{Barroso2007Case, Lo:2014:Towards}, often more than 30\% of their peak power usage. However, for the processing component, the idle power is significantly lower. 
Finally, while individual servers may be fully utilized, the datacenter-level utilization ranges from 30\% to 60\% even for the state-of-the-art datacenters~\cite{Tirmazi:2020:Borg}.

\vspace{0.05cm}
\noindent\textbf{-- Energy and Carbon Footprint Estimates.} The energy consumption and carbon footprint at the server level also depends on the power supplies used, the number of hard drives, the memory size, and the chassis, among other components. 
We use the inventory information from MIT's academic clusters, including the Bates Research and Engineering Center~\cite{MIT:2024:Bates} and the hydro-powered Massachusetts Green High Performance Computing Center (MGHPCC)~\cite{MIT:2024:MGHPCC}.
We only have information on the processor component of the server. 
As a result, our current embodied and operational carbon estimates only account for the processing component.

The embodied carbon for the processor is estimated using a research version of the integrated circuit module of PAIA~\cite{Olivetti:2012:PAIA}, which uses information on the technology node (such as 7nm or 28nm), CPU package area, die size, and the fabrication location. 
The official Intel and AMD websites often provide data on the technology node and CPU package area but do not provide information on the die size. 
We obtained the die sizes using data from  the TechPowerUp~\cite{TechPowerUp:2024:CPU}, CPU-World~\cite{CPUWorld:2024:CPU}, X86 CPU's Guide~\cite{CPUGuide:2024:CPU}, and WikiChip~\cite{WikiChip:2024:CPU} websites. 
Since these are not official websites, we ensured that the die sizes across websites were consistent; we only used processors with consistent information across at least two websites.
For AMD processors, we used carbon intensity values for Taiwan, which is 495\CI~\cite{Taiwan:2023:CI}. For Intel, we assume fabrication in Hillsboro, Oregon, with a carbon intensity of 357\CI~\cite{electricity-map}.

For the operational carbon estimates, we assume a server consumes its rated Thermal Design Power (TDP) at 100\% utilization, with a linear increase in power between the extremes.   For the carbon intensity, unless otherwise specified, we assume the datacenter is situated in Sweden and has a carbon intensity of 20\CI~\cite{electricity-map}. In~\autoref{sec:illustration}, we vary the carbon intensity for analysis in embodied- and operational-dominant regions. 

\vspace{0.05cm}
\noindent\textbf{-- Performance Benchmarks.} We use three benchmarks to get performance scores for the processors: Multithread Ratings for CPUs by PassMark~\cite{PassMark:2024:CPU}, HEPScore~\cite{HEPScore:2024:CPU}, and SPEC CPU2017 Floating Point Speed~\cite{SPECScore:2024:CPU}. 
However, not all the benchmarks profile each processor, narrowing the set of processors used in our analysis.  

\subsection{Carbon-Aware Scheduling Metrics}
\label{sec:illustration}
This section defines three different metrics that can be used to evaluate carbon-aware scheduling and job placement.

\smallskip
\noindent
\textbf{1 -- Software Carbon Intensity (\SCI)} was introduced by the Green Software Foundation~\cite{GSF:2021:SCI}; it quantifies the rate of total carbon emissions per functional unit \textbf{\texttt{R}}. 
The functional unit here can be an API call, ML training, or large language model (LLM) inference.

The carbon emissions for a given job include both the operational carbon emissions (denoted by \texttt{O}) for running the job on the server and the embodied carbon emissions (denoted by \texttt{M}) for the functional unit representing the job. 
\SCI is expressed as, \\
\vspace{-0.3cm}
\[ \SCI = (\texttt{O} + \texttt{M})\ \text{per}\ \texttt{R}.\]
\[ \SCI = (\texttt{(E * I)} + \texttt{M})\ \text{per}\ \texttt{R}.\]
Here, \texttt{E} is the energy consumption in kilowatt-hours of the job over a given time window. 
This includes a portion of the idle power for the server assigned to the job and dynamic power due to the job's resource usage. \texttt{I} is the carbon intensity of electricity in grams of carbon dioxide equivalent per kilowatt-hour (\CI) for the region where the server consumes electricity.

\SCI only accounts for the embodied carbon ($\texttt{M}$) of the active server running the job with its value is computed as, 
\begin{equation}
\label{eq:proportion}
\texttt{M} = \texttt{TE} \times \frac{\texttt{T} \times \texttt{RR}}{\texttt{EL} \times \texttt{TR}}.
\end{equation}
Here, \texttt{TE} is the total embodied emissions, \texttt{EL} is the expected lifespan, and \texttt{TR} is the server's total resources for the server running the job. 
\texttt{T} is the time duration and \texttt{RR} is the resource reserved for the job. 
Note that SCI ignores the embodied and operational emissions of the idle servers in the datacenter, as it focuses on accounting for a given job's carbon footprint (see SCI specifications for details~\cite{GSF:2021:SCI}).

\vspace{0.1cm} 
\noindent
\textbf{2 -- Total Software Carbon Intensity (\TSCI)}  
extends \textbf{\SCI} by considering the embodied carbon emissions at the infrastructure scale, a potential solution to accurately account for total emissions. This metric assigns a portion of the total embodied emissions for the infrastructure to each running job instead of just considering the embodied emissions for the server that runs the job.

In extending \SCI, which already accounts for the embodied carbon for the server that runs the job, we must add a fraction of the infrastructure-level embodied carbon emissions proportional to the resources reserved and allotted time for the job. 
This can be simplified by taking the datacenter's total embodied carbon and assigning a portion of that to the job. In this case, the total embodied carbon for the job (\texttt{tM}) is computed as, 
\[ \texttt{tM} = \texttt{M} + \texttt{M}_\text{idle-infra}. \]
The value of \texttt{M}$_\text{idle-infra}$ is calculated as the sum of \texttt{M} for all the idle servers using~\autoref{eq:proportion}, i.e., each idle server's embodied carbon is also proportionally assigned to the job. Similarly, 
the operational carbon from the base power consumption of idle servers also contributes to \texttt{tO}, the total operational carbon footprint, 
\[ \texttt{tO} = \texttt{O} + \texttt{O}_\text{idle-infra}. \]
Finally, the total software carbon intensity can be computed as, 
\[ \TSCI = (\texttt{tO} + \texttt{tM})\ \text{per}\ \texttt{R}.\]
To show how to calculate this value, consider an example where there is a datacenter with two servers A and B, with embodied carbon values of 400\texttt{g.CO$_\text{2}$} and 50\texttt{g.CO$_\text{2}$} and an expected lifetime of 10 years and 5 years, respectively. 
Assume server A has 40 cores, and server B has 10 cores. 
Suppose a job \texttt{J$_1$} that runs for one year and uses 10 cores is scheduled on server B. Another job \texttt{J$_2$} runs on server A and uses 10 cores; the value of embodied carbon attributed to \texttt{J$_1$} will be computed as, 
\[ \texttt{tM} =  10\texttt{g.CO$_\text{2}$} + \underbrace{\frac{400\texttt{g.CO$_\text{2}$} \times 1 yr}{10 yrs}}_\text{time fraction} \times \underbrace{\frac{30 cores}{40 cores}}_\text{idle fraction} \times \underbrace{\frac{10 cores}{20 cores}}_\text{usage fraction}, \]
\[ = 25\texttt{g.CO$_\text{2}$}.\]
In the fraction above, the time, idle, and usage fractions are the amortization terms that amortize the embodied carbon of the idle infrastructure over time (1 out of 10 years), idle resources (30 out of the 40 cores in the remaining infrastructure are idle), and usage (job uses 10 of the total 20 cores used). Since operational carbon emission rates are instantaneous, the value of \texttt{tO} can also be calculated using the same method, except for the time fraction component. 

\smallskip
\noindent
\textbf{3 -- Operational Software Carbon Intensity (\OSCI)} metric ignores the embodied carbon emissions for all the servers. It makes scheduling decisions based on the operational carbon emissions of running a given job. \OSCI is expressed as, \\
    \vspace{-0.3cm}
    \[ \OSCI = (\texttt{E * I})\  \text{per}\ \texttt{R}.\]
This metric can include a portion of the base power from the idle servers to incentivize turning off servers when they are not needed. 
However, for the current purpose, we keep it simple and only account for the energy used by the server running the job.

\smallskip
\noindent
\textbf{Computing \SCI, \TSCI, and \OSCI in Practice} involves different degrees of challenges. First, \OSCI is a subset of the other two metrics and is the simplest to calculate, as the operating power of a job can be estimated through offline profiling.  \SCI requires embodied carbon estimates for all the servers in a datacenter, which can be difficult to obtain in practice.
Note that embodied carbon estimates also tend to have a high degree of uncertainty~\cite{Bhagavathula:2024:EmbodiedUncertainty, Alcaraz:2018:StreamLinedLCA, Olivetti:2012:PAIA}. As a result, the uncertainty in the embodied carbon estimates can propagate and affect the scheduling outcomes unpredictably.

Finally, calculating \TSCI and tracking it over time is complex and necessitates comprehensive datacenter-level information, encompassing all hardware components and active jobs, including their resource reservations and expected runtime. 
Also, as the jobs arrive and leave, the idle fraction of the infrastructure will change, resulting in a time-varying value of \TSCI. 
While cloud operators have access to this data, calculating \TSCI demands sophisticated data collection infrastructure and precise online attribution, which entail considerable cost and carbon overheads. Such detailed information is generally not accessible to end users in many contexts, such as public cloud environments, prohibiting them from computing their carbon footprint if the cloud providers do not share this information. 
Therefore, we do not envision this metric being used in practice; instead, we include it for completeness of the metrics and show that a more straightforward metric of \OSCI can achieve the same scheduling outcomes. 

Finally, incorporating information on operational and embodied carbon estimates into scheduling decisions depends on the scheduler being used. For example, in Slurm, nodes can be assigned arbitrary weights that determine their priority for scheduling; these weights can be set to the value of the metric of choice, such as \OSCI.  To compute \OSCI values, Slurm's or any other resource manager's energy monitoring tool can be easily augmented to report operational emissions with little overhead.

\begin{table}[t]
    \footnotesize
    \centering  
    \vspace{0.15cm}
    \captionof{table}{\textbf{Specifications of servers in our illustrative example.}}
    \vspace{-0.3cm}
        \begin{tabular}{|l|c|c|}
            \hline
            \textbf{} & \textbf{$S_A$} & \textbf{$S_B$}  \\ \hline\hline 
            \textbf{Processor} & \texttt{Xeon E-2286G} & \texttt{Xeon Gold 6538N} \\\hline 
            \textbf{Release Date} & 05/29/2019 & 12/14/2023 \\\hline 
            \textbf{PassMark Score} & 14020 & 44895 \\\hline
            \textbf{TDP (W)} & 90 & 205 \\\hline
            \textbf{Technology Node} & 14nm & 10nm \\\hline 
            \textbf{Embodied Carbon (Kg.CO2)} & 8.04 & 101.89 \\\hline\hline
        \end{tabular}
        \vspace{-0.35cm}
    \label{tab:illustrative-example-setup}
\end{table}

\begin{table}[t]
    \footnotesize
    \centering  
    \vspace{0.15cm}
    \captionof{table}{\textbf{Values of \SCI, \TSCI, \OSCI for \textbf{$S_A$} and \textbf{$S_B$} for job placement in \C per \texttt{Score-Yr}. We also report the total cluster carbon footprint for each metric.}}
    \vspace{-0.3cm}
        \begin{tabular}{|l|c|c|c|}
            \hline
            \textbf{Metric} & \multicolumn{2}{|c|}{\textbf{Scheduling/Placement}} & \textbf{Accounting} \\ \hline \hline 
             & \textbf{$S_A$} & \textbf{$S_B$} & Cluster Carbon Footprint\\ \hline
            \textbf{\SCI} & 0.11 + 0.83 = \textbf{0.94} & 0.45 + 0.56 = 1.01 & (0.11 + 0.45) + 0.83 = 1.39 \\ \hline
            \textbf{\TSCI} & 0.94 + 0.45 = 1.39 & 1.01 + 0.11 = \textbf{1.12} & (0.11 + 0.45) + 0.56 = 1.12 \\ \hline
            \textbf{\OSCI} & 0.83 & \textbf{0.56} & (0.11 + 0.45) + 0.56 = 1.12 \\ \hline \hline 
        \end{tabular}
        \vspace{-0.35cm}
    \label{tab:illustrative-example}
\end{table}

\subsection{An Illustrative Example}
\label{sec:illustration}
We first use an illustrative example to demonstrate the \emph{sunk carbon fallacy}. 
Consider a small datacenter with two servers powered by two processors from Intel: \texttt{Xeon E-2286G} and \texttt{Xeon Gold 6538N}.
We refer to these two servers as \SA and \SB, respectively. 
\autoref{tab:illustrative-example-setup} provides the detailed specifications for the two servers, including processor model, their release dates, PassMark scores, embodied carbon estimates, and TDP values. 

\begin{figure}[t]
  \begin{center}
    \includegraphics[width=\columnwidth]{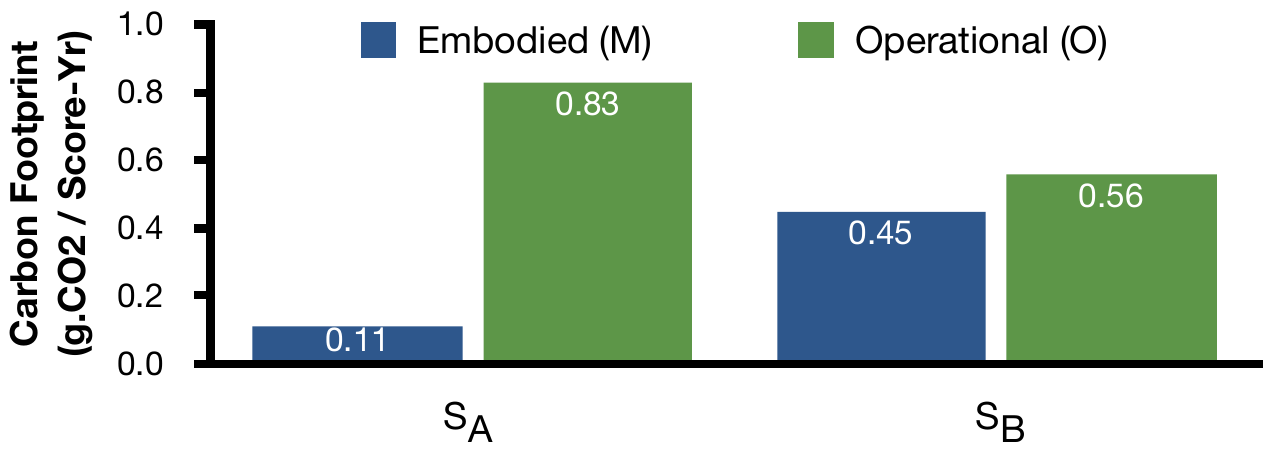}
  \end{center}
  \vspace{-0.4cm}
  \caption{The normalized embodied and operational carbon footprint (\C) per \texttt{Score-Yr} assuming the datacenter is located in Sweden and the electricity has a carbon intensity of 14\CI~\cite{electricity-map}. The servers have a lifetime of 5 years.}
    \vspace{-0.5cm}
    \label{fig:servers-example}
\end{figure}

\autoref{fig:servers-example} shows the operational and embodied carbon emissions normalized to the PassMark score and the server's expected lifetime for the two servers in our illustrative datacenter. 
Intuitively, an operational carbon value of 0.56 means that getting a performance of 1 score for one year using \SA will incur operational emissions of 0.56\C. 
Note that our illustrative example maps to an increasingly common real-world scenario, where a newer server (\SB) manufactured using recent technology, 10nm in this case, has 4.09$\times$ higher embodied carbon footprint than an older server (\SA) manufactured with 14nm technology. 
However, the energy efficiency gains over the last few years mean that \SB is significantly more energy-efficient (consumes 32.5\% less energy) than \SA. 

\setlength{\intextsep}{0pt}%
\setlength{\columnsep}{8pt}%

\setlength{\tabcolsep}{4pt}
\begin{table}[t]
    \footnotesize
    \centering  
    \vspace{0.15cm}
    \captionof{table}{\textbf{Additional example scenarios that may lead to sunk carbon fallacy, i.e., an inefficient server with low \SCI value is used before an efficient server with high \SCI value. Values of carbon emissions are in \C per \texttt{Score-Yr}. }}
    \vspace{-0.3cm}
        \begin{tabular}{|c|c|c|}
            \hline
             \multicolumn{2}{|c|}{\textbf{Server Pairs}} & \textbf{Additional Details} \\ \hline \hline 
             Xeon E-2486 & EPYC 9334 & New Xeon server (12/14/2023, 10nm)\\ 
             0.08 + 0.47 = \textbf{0.55} & 0.23 + 0.39 = 0.62 & vs. old EPYC server (11/10/2022, 5nm).  \\ \hline
             Ryzen 5965WX & Xeon W9-3495 & Older Ryzen server (03/08/2022, 5nm)\\ 
             0.15 + 0.51 = \textbf{0.66} & 0.25 + 0.46 = 0.71 & vs. New Xeon server (02/15/2023, 10nm).  \\ \hline\hline 
        \end{tabular}
        \vspace{-0.45cm}
    \label{tab:additional-examples}
\end{table}

\smallskip
\noindent
\textbf{1 -- Analyzing Scheduling Outcomes.}
\autoref{tab:illustrative-example} shows the carbon footprint values used to choose one of the servers for job placement. We also report the total lifecycle emissions of the datacenter for the duration of the job, including the embodied carbon footprint of all the servers and the operational carbon footprint of the active servers. 
The values for the server with the lowest metric are highlighted in bold; the server with the lowest value is chosen to run the job. We compute the datacenter-level carbon footprint as the sum of embodied carbon for all the servers (the sunk cost) and operational carbon for the server running the job (the marginal or additional cost). 
As shown, in prioritizing the sum of embodied and operational, \SCI chooses a highly energy-inefficient server with a low \SCI value due to a small embodied carbon value. 
While this placement is preferable based on the \SCI metric, it leads to a 24.10\% higher carbon footprint for the cluster.
On the other hand, the placement choices of \TSCI and \OSCI align and lead to the minimum value of the cluster-level carbon footprint as both minimize the additional emissions to obtain the desired performance.

In this example, we used the classic case of a new efficient server with high embodied carbon against an old energy-inefficient server with low embodied carbon, primarily due to the technology node difference. 
However, this discrepancy of an energy-inefficient server having a lower \SCI value than an energy-efficient server can also occur in other scenarios. For example, as shown in~\autoref{tab:additional-examples}, the new \texttt{Xeon E-2486} server uses a 10nm technology node and has a smaller embodied carbon than its counterpart \texttt{EPYC 9334} server. 
The energy efficiency gains and performance improvement for \texttt{EPYC 9334} from advanced manufacturing are not enough to outweigh the increase in embodied carbon, leading to its higher \SCI value.
Surprisingly, a similar discrepancy occurs between \texttt{Ryzen Threadripper 5965WX} and \texttt{Xeon W9-3495} as the former has a lower embodied carbon footprint due to a smaller die despite using a 5nm technology node as compared to 10nm for the latter. 
We hand-picked these examples to demonstrate the existence of the \emph{sunk carbon fallacy} beyond the classic old vs. new example. While these servers are not like-for-like replacements for each other, they can still be available in a given datacenter or a cloud platform, leading to the choice of an inefficient server over an energy-efficient one.

\begin{figure*}[t]
    \vspace{-0.5em}
    \minipage{0.24\textwidth}
    \includegraphics[width=\linewidth]{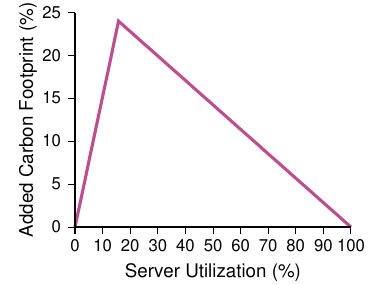}\vspace{-1em}
    \caption{Utilization Impact}\label{fig:utilization}
    \endminipage\hfill
    \minipage{0.24\textwidth}
    \includegraphics[width=\linewidth]{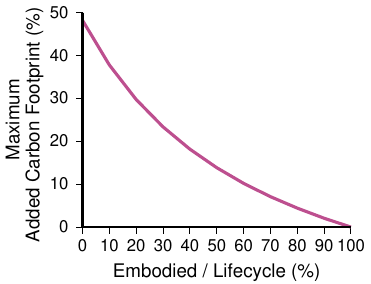}\vspace{-1em}
    \caption{Op. Carbon Impact}\label{fig:emb_frac}
    \endminipage\hfill
    \minipage{0.24\textwidth}
    \includegraphics[width=\linewidth]{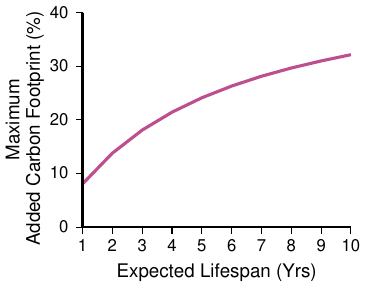}\vspace{-1em}
     \caption{Lifespan Impact}\label{fig:lifecycle}
    \endminipage\hfill  
    \minipage{0.27\textwidth}
    \includegraphics[width=\linewidth]{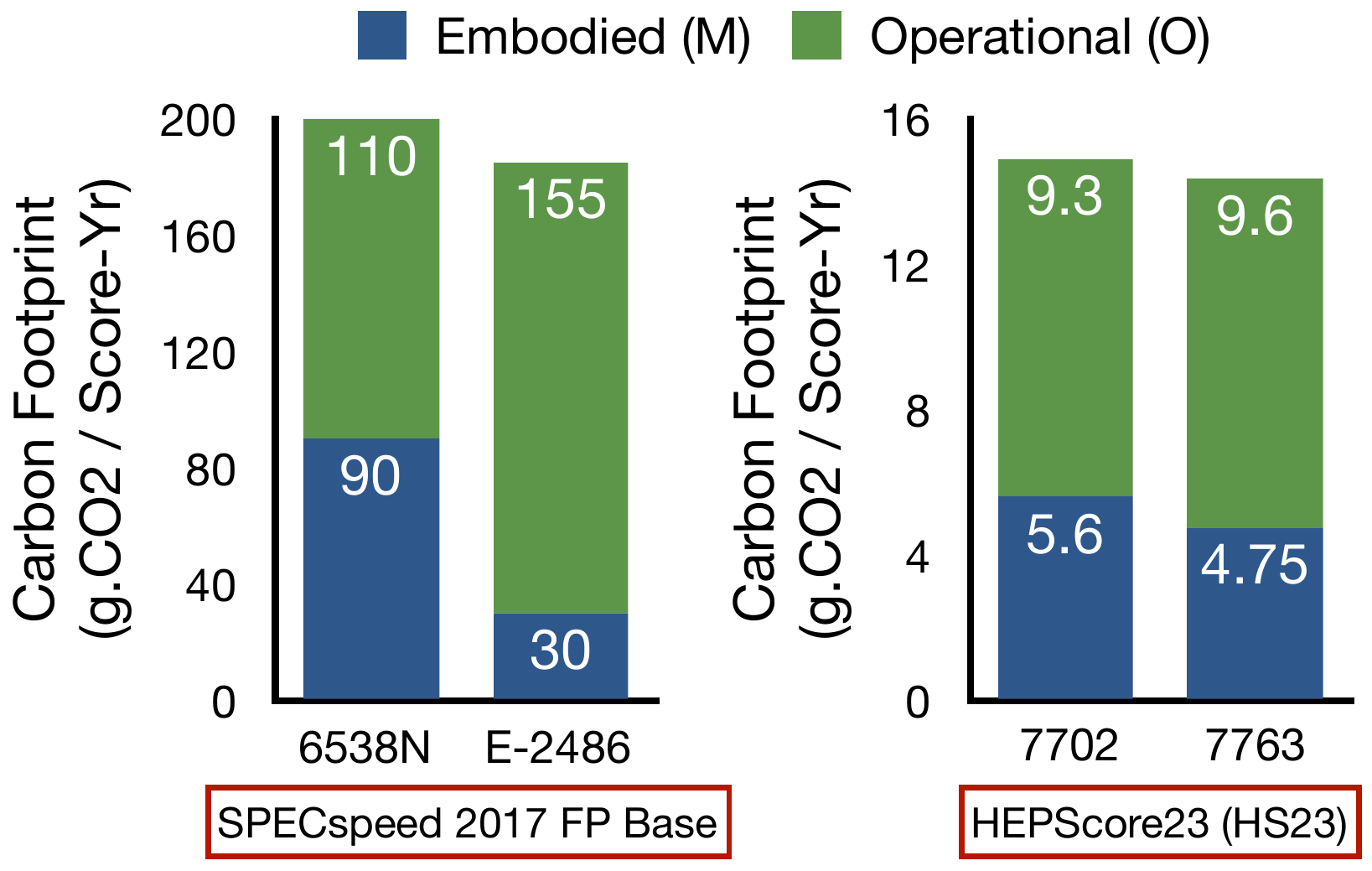}\vspace{-0.2cm}
     \caption{Benchmark Impact}\label{fig:score}
    \endminipage
    \vspace{-0.3cm}
\end{figure*}

\smallskip
\noindent
\textbf{2 -- Effect of Datacenter Utilization.} 
Our illustrative example shows how differences in server characteristics lead to suboptimal scheduling outcomes. 
We next analyze the effect of utilization on the increase in the system-level carbon footprint when using \SCI. 
The server \SA has 12 logical cores (6 physical cores and 2 threads per core), where each logical core provides a performance score of 1168. 
The server \SB has 64 logical cores (32 physical cores and 2 threads per core), where each logical core provides a performance score of 701.
Each job uses one logical core on \SA. 
On \SB, each job uses 2 logical cores to get a 1402 performance score (closer to 1168 score for \SA). We have a total of 44 cores of similar performance. 

\vspace{0.1cm}
\autoref{fig:utilization} shows the increase in the system-level carbon footprint when jobs are scheduled using \SCI compared to scheduling based on \TSCI or \OSCI.
If no server in the datacenter is being used or all servers are being used, all the metrics yield the same outcome. 
However, when the utilization levels are between 0 and 100\%, the set of servers selected to run the job matters. 
The peak discrepancy happens when only 12 cores are needed to run the jobs (at 27.3\% utilization) and decreases afterwards.
The exact magnitude of the peak and the utilization at which this manifests will change based on the set of servers, their base power values, and the granularity at which the jobs can be scheduled. In~\autoref{sec:case-study}, we provide the same results for the academic datacenter we study.

\smallskip
\noindent
\textbf{3 -- Effect of Operational Carbon Intensity.}
In our current setup, embodied carbon accounts for 11.7\% and 44.5\% of the lifecycle emissions for \SA and \SB, respectively. The average value across servers is 28.1\%. 
We scale our normalized operational carbon footprint to generate values such that embodied carbon accounts for 10\% to 90\% of the lifecycle emissions and study its impact on the increase in system-level footprint due to \SCI. 
\autoref{fig:emb_frac} shows the maximum value of the added carbon footprint due to the sunk carbon fallacy as the embodied carbon footprint accounts for an increasing fraction of the lifecycle emissions. 
At 0\%, only the operational efficiency matters and the use of \SA results in a 48\% increase in the system-level carbon footprint. 
At the other extreme of 100\%, the operational carbon is 0 and the choice of server does not matter. 

\vspace{0.1cm}
We note that operational carbon emissions dominate despite our use of Sweden for the datacenter location, one of the greenest regions in the world with only 20\CI. 
This is because our embodied carbon estimates only include the processor component that contributes a small fraction of the overall server-level carbon footprint.
However, the TDP value of the processor component accounts for most of the server-level power and operational carbon footprint. 
If server-level embodied carbon values are used, the carbon intensity values at which embodied carbon accounts for a given percentage of lifecycle emissions will be higher.

\setlength{\tabcolsep}{3.0pt}
\begin{table*}[t]
    \centering
    \footnotesize
    \caption{List of servers and their specifications for the case study datacenter. The embodied carbon values only account for the processor. The server life is assumed to be 5 years if it is less than 5 years old; otherwise, its embodied carbon is amortized over all the years since purchase. The operational carbon values are for five years and computed with a carbon intensity of 10 \CI (chosen such that embodied carbon accounts for 20\% of the lifecycle emissions).}
    \vspace{-0.35cm}
    \begin{tabular}{l|ll|cc|cccc|c|ccc}
        \toprule
        \textbf{Processor} & \textbf{Purchase} & \textbf{Server} & \textbf{Technology} & \textbf{Embodied Carbon} & \multicolumn{4}{c|}{\textbf{Performance \& Power}} & \textbf{Operational Carbon} & \multicolumn{3}{|c}{\textbf{Carbon (\C\texttt{/Score-Yr}})} \\ \cline{6-9}
        
        \rule{0pt}{2.5ex} & \textbf{Year} & \textbf{Count} & \textbf{Node} & \textbf{(\textbf{\texttt{KgCO$_\text{2}$}})}   & \textbf{PassMark} & \textbf{TDP (W)} & \textbf{Cores} & \textbf{Threads} &  \textbf{(\textbf{\texttt{KgCO$_\text{2}$}})}  & \texttt{\textbf{M}} & \texttt{\textbf{O}} & \textbf{\SCI} \\
        \midrule
Xeon-Silver-4216 & 2020 & 59 & 14 & 24.15 & 20613 & 100 & 16 & 32 & 43.80 & 0.234 & 0.425 & 0.659 \\
Xeon-Silver-4116 & 2019 & 109 & 14 & 21.18 & 14660 & 85 & 12 & 24 & 37.23 & 0.289 & 0.508 & 0.797 \\
Xeon-E5-2640v4 & 2016 & 54 & 14 & 19.08 & 12472 & 90 & 10 & 20 & 39.42 & 0.194 & 0.632 & 0.826 \\
Xeon-E5-2640v3 & 2015 & 65 & 22 & 19.36 & 11118 & 90 & 8 & 16 & 39.42 & 0.183 & 0.709 & 0.892 \\
Xeon-E5-2650v2 & 2014 & 36 & 22 & 09.44 & 9866 & 95 & 8 & 16 & 41.61 & 0.096 & 0.844 & 0.939 \\
Xeon-E5-2620-v4 & 2017 & 30 & 14 & 13.47 & 9193 & 85 & 8 & 16 & 37.23 & 0.209 & 0.810 & 1.019 \\
Xeon-Gold-6326 & 2021 & 68 & 10 & 101.0 & 35270 & 185 & 16 & 32 & 81.03 & 0.573 & 0.459 & 1.032 \\
Xeon-E5640 & 2012 & 47 & 32 & 11.39 & 3782 & 80 & 4 & 8 & 35.04 & 0.251 & 1.853 & 2.104 \\
Xeon-E5620 & 2010 & 52 & 32 & 12.71 & 3590 & 80 & 4 & 8 & 35.04 & 0.253 & 1.952 & 2.205 \\
Xeon-E5-2609-v2 & 2014 & 22 & 22 & 10.49 & 3369 & 80 & 4 & 4 & 35.04 & 0.312 & 2.080 & 2.392 \\
Xeon-X5647 & 2012 & 82 & 32 & 13.45 & 4441 & 130 & 4 & 8 & 56.94 & 0.253 & 2.564 & 2.818 \\
Xeon-E5520 & 2010 & 25 & 45 & 12.12 & 2524 & 80 & 4 & 8 & 35.04 & 0.343 & 2.777 & 3.120 \\
Xeon-E5410 & 2008 & 43 & 65 & 11.75 & 2007 & 80 & 4 & 4 & 35.04 & 0.365 & 3.492 & 3.857 \\
Xeon-E5335 & 2007 & 28 & 65 & 13.45 & 1549 & 80 & 4 & 4 & 35.04 & 0.542 & 4.524 & 5.066 \\
Xeon-E5310 & 2007 & 20 & 65 & 14.19 & 1306 & 80 & 4 & 4 & 35.04 & 0.639 & 5.366 & 6.005 \\

\midrule
\textbf{Total} & -- & \textbf{740} & -- & \textbf{17632.71} & \textbf{8261198} & \textbf{74045} & \textbf{6204} & \textbf{11956} & -- & -- & -- & -- \\
        \bottomrule 
    \end{tabular}
    \label{tab:case-study}
    \vspace{-0.35cm}
\end{table*}

\smallskip
\noindent
\textbf{4 -- Effect of Server's Expected Lifetime.} The expected lifespan of servers has a similar impact on the added carbon footprint at the system level. ~\autoref{fig:lifecycle} shows the maximum added carbon footprint at the system-level as the server's embodied carbon is amortized over a longer period.
As the expected lifespan increases, the amortized embodied carbon per year decreases, and its fraction of the lifecycle carbon footprint decreases. As shown in~\autoref{fig:emb_frac}, lower embodied values result in a higher system-level carbon footprint under \SCI, magnifying the impact of the \emph{sunk carbon fallacy}.

\smallskip
\noindent
\textbf{5 -- Effect of Performance Metric.} Our results thus far have used PassMark scores. However, our observation is agnostic to any particular benchmarking method. ~\autoref{fig:score} shows that the conditions required for the \emph{sunk carbon fallacy}, i.e., a server with low \SCI is inefficient, manifest across different benchmarks. The servers we use in our examples changed, as we did not have SPEC and HS26 scores for the servers in the illustrative example. 
While the combination of servers that manifest the \emph{sunk carbon fallacy} may change, the effect should be present in all performance benchmarks.

\subsection{Generalization of Outcomes}
\label{sec:general}
We next review if our observations hold for all scenarios of hardware choices (concerning their embodied and operational carbon ratios). 
Let us assume there are $\text{N}$ servers in a datacenter, and we need $\text{k}$ servers at a time. 
Let \( M_i \), \( O_i \) be the embodied and operational carbon costs of server \( i \). 
Let \( Z_i = M_i + O_i \) be the total carbon emissions for a functional unit or over the server's lifetime. 

The \SCI and \OSCI strategies can be written as:
\begin{align*}
    \SCI &= \{ i \mid Z_i \text{ are the } k \text{ smallest values of } Z \}, \\
    \OSCI &= \{ i \mid O_i \text{ are the } k \text{ smallest values of } O \}.
\end{align*}

If k is zero or equals N, both strategies yield the same set of servers. 
\OSCI directly minimizes \( \sum_{i \in \OSCI} O_i \), the operational carbon, which is the only cost that can be reduced post-purchase. Since \SCI might include servers with a lower lifecycle cost \( Z_i \) but potentially higher \( O_i \), \OSCI can yield a lower total carbon when both embodied and operating carbon are considered together. Therefore:
\[
\sum_{i \in \OSCI} O_i \leq \sum_{i \in \SCI} O_i.
\]
Given the above inequality and considering the total carbon footprint \( Z_i = M_i + O_i \) across embodied and operation al phases across all the servers in the datacenter, the choice of \OSCI ensures the minimum total carbon across purchase and operation. 

Extending our example to show that operational carbon emissions yield the lowest carbon footprint even when jobs arrive over time is straightforward. However, doing so is outside the scope of this vision paper and a subject of future work. 

\section{Case Study: An Academic Datacenter}
\label{sec:case-study}
In the previous section, we used two simple servers to illustrate the effect of different metrics and how various server specifications, datacenter characteristics, and accounting considerations influence the \emph{sunk carbon fallacy}. 
As a case study, we use an MIT academic datacenter that runs scientific computing workloads~\cite{MIT:2024:Bates, MIT:2024:MGHPCC}.

Our case study demonstrates that the \emph{sunk carbon fallacy} is not an artifact of our illustrative example; it manifests itself in real-world datacenters with an arbitrary set of servers. Our findings and analysis predicate the assumption that carbon-aware scheduling aims to reduce the total cluster-level carbon footprint (embodied and operational) of running a set of jobs on a given set of servers. 

\vspace{0.05cm}
\noindent
\textbf{1 -- Case Study Setup.}
We use the setup described in~\autoref{sec:setup}, except where noted below. 
\autoref{tab:case-study} shows the detailed specifications of the servers in our real-world case study datacenter. 
Our datacenter inventory contains 15 different processor types across 740 servers. 
The average age of a server is 9.5 years: the oldest servers (\texttt{E5310} and \texttt{E5335}) are 17 years old, the newest servers (\texttt{Gold-6326}) are just 3 years old, and only 236 out of 740 servers (31.9\%) less than five years old. 
All processors are Intel-manufactured using technology nodes ranging from 64nm to 10nm. 
The embodied carbon of processors in servers ranges from 9.44 \texttt{KgCO$_\text{2}$} to 101.0 \texttt{KgCO$_\text{2}$} with a total of 17,633 \texttt{KgCO$_\text{2}$} embodied carbon from processing component of servers. 
The PassMark score (multi-threaded rating) for the processors has a wide range: 1306 for \texttt{E5310} (the oldest processor) to 35,270 for \texttt{Gold-6326} (the newest processor). The TDP values are also at the extreme ends for the two processors, with 80W for \texttt{E5310} and 185W for the \texttt{Gold-6326}. 

We take a server's expected lifespan to be 5 years, which is generally true for modern datacenters. However, servers are typically kept operational in academic datacenters as purchasing new hardware necessitates considerations beyond performance and operating costs. 
We use two accounting approaches to amortize embodied carbon over a server's lifespan: first, the embodied carbon of servers older than five years is set to 0, and second, embodied carbon is amortized over the years they have been operational. 
In \autoref{tab:case-study}, the normalized values of operational carbon (\texttt{O}), embodied carbon (\texttt{M}), and software carbon intensity (\SCI) use the latter approach. 
We choose the latter as setting embodied carbon to 0 for older servers artificially inflates the \emph{sunk carbon fallacy}. We later show the impact of both approaches on the added carbon footprint.

\setlength{\intextsep}{0pt}%
\setlength{\columnsep}{8pt}%
\begin{figure}[t]
  \begin{center}
    \includegraphics[width=\columnwidth]{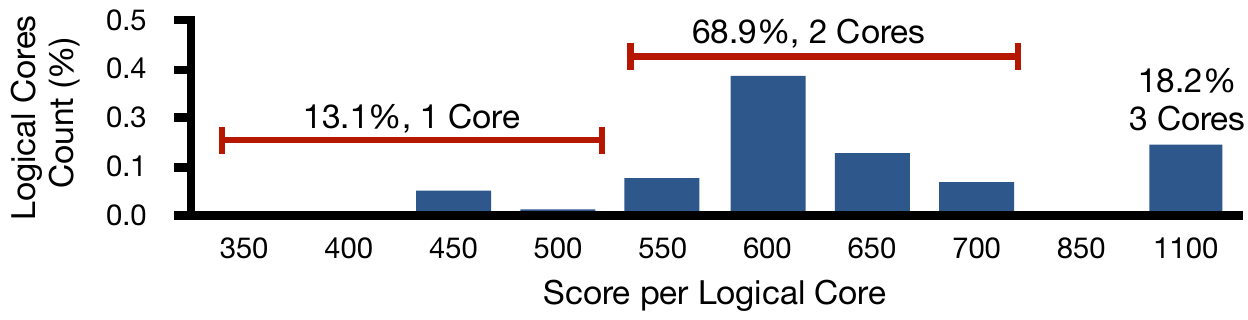}
  \end{center}
  \vspace{-0.55cm}
  \caption{Normalized logical cores of similar performance.}
    \vspace{-0.7cm}
    \label{fig:core_distribution}
\end{figure}

We use job traces from a year-long trace from 2016 consisting of 14M jobs from the MGHPCC cluster used in prior work~\cite{Ambati:2020:WaitingGame, Ambati:2021:GoodThings}. The trace provides information on the job submission time, job end time, the requested number of cores, and the amount of memory requested (not used in our analysis). 
To get core allocations that provide roughly similar performance across these heterogeneous machines, we normalize the machines with thread count and categorize threads into three virtual core categories, as shown in~\autoref{fig:core_distribution}: \textbf{VC1} has 13.1\% of threads with 250--500 score, \textbf{VC2} has 68.9\% of threads with 550--700 score (2$\times$\textbf{VC1}), and \textbf{VC3} has 18.2\% of threads with (3$\times$\textbf{VC1}).
Since the biggest server in our case study datacenter has 32 threads, we filter all the jobs that require more than 32 cores.

\vspace{0.05cm}
\noindent
\textbf{2 -- Case Study Findings.}
\autoref{tab:case-study} shows \SCI values for the servers sorted in ascending order of \SCI, which is the energy-efficient ordering. 
For example, based on \SCI metric, \texttt{Xeon-E5-2620-v4} will be chosen before \texttt{Xeon-Gold-6326} despite the former having 1.37$\times$ higher carbon footprint. 
\texttt{Xeon-Gold-6326} is the second most efficient server, but it is 7th on the \SCI ranking. 
The order is also suboptimal in other cases, such as \texttt{Xeon-E5-2620-v4} and \texttt{Xeon-E5-2650-v2}, where an even less efficient server gets picked due to its lower embodied carbon. 
If the embodied carbon of servers over five years old is set to 0, the order is impacted even further. 
The three most efficient servers, \texttt{Silver-4216}, \texttt{Gold-6326}, and \texttt{Silver-4116}, will be ranked 2nd, 7th, and 4th, respectively.

While these ranking changes may seem minor, they can result in significant added carbon at the datacenter level when using \SCI. 
To assess the datacenter-level impact, we compute the added carbon under \SCI and \OSCI. 
To do that, we place the jobs on the servers based on their submission time. 
We do not replay the job trace and perform one-time job placement, akin to placing long-running jobs that never finish. 
Each job requires a certain number of virtual cores, and multiple jobs can be allocated to one server, which avoids stranded resources.
A more realistic replay of job trace and placement is outside the scope of this work.

\begin{figure}[t]
    \minipage{0.22\textwidth}
    \includegraphics[width=\linewidth]{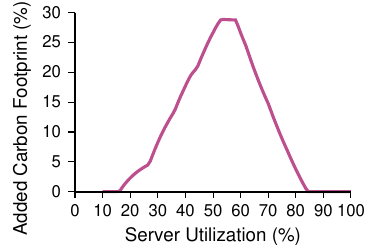}\vspace{-0.4cm}
    \caption{Embodied amortized across the lifespan.}\label{fig:non_zero_approach}
    \endminipage
    \hfill
    \minipage{0.22\textwidth}
    \includegraphics[width=\linewidth]{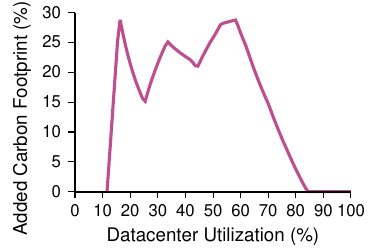}\vspace{-0.4cm}
     \caption{Embodied amortized during the first 5 years.}\label{fig:zero_approach}
    \endminipage
    \vspace{-0.65cm}
\end{figure}

\autoref{fig:non_zero_approach} and \autoref{fig:zero_approach} show the added carbon due to \SCI for the two approaches to amortize the embodied carbon for older servers.
In both cases, using \SCI results in nearly a 30\% increase in carbon footprint for the datacenter due to the use of energy-inefficient servers. 
The added carbon for the first amortization approach is more than 5\% when the datacenter utilization is between 27\% and 78\%, a range in which almost all datacenters operate.
The second approach has an even higher added carbon cost (typically above 10\%) across a broader utilization range of 13\% to 80\%. 
This demonstrates that even a slight change in the order of servers can significantly impact datacenter-level carbon. 
Furthermore, this result also shows how \SCI is susceptible to an arbitrary setting of the expected lifespan. 
Finally, the cluster utilization in our job trace ranges from 40-80\%; thus, using \SCI will incur at $>$15\% higher carbon footprint. 
Note that the first approach of amortizing over increasing periods beyond 5 years leads to double-counting of embodied carbon, as all embodied carbon would have been accounted for in 5 years.

\vspace{-0.2cm}
\section{Implications and Conclusion}
\label{sec:path-forward}
\vspace{-0.05cm}
Next, we discuss the implications of using the three carbon-based metrics to schedule jobs on on-premise and cloud datacenters.

\textbf{\SCI} quantifies the total carbon footprint of a functional unit, incorporating both operational and embodied emissions. 
\SCI is an intuitive and comprehensive metric, but it is not well-suited for some decisions in sustainable computing.
The metric implicitly requires that, for a server to be preferred over a reference, any increase in the server's embodied carbon footprint must reduce operational emissions by the same or a higher margin. 
However, since embodied carbon and operational carbon occur at different timescales, an arbitrary setting of the server's expected lifespan and embodied carbon accounting approach can perturb the embodied-to-operational carbon ratio. 
For example, as shown in Figures~\ref{fig:lifecycle}--\ref{fig:zero_approach}, different accounting approaches for embodied carbon and varying expected lifespans led to different operational carbon.

One key aspect of \SCI is that it incentivizes using older hardware, which may have much lower embodied carbon per score (due to older and less energy-intense technology) than the newer servers with typically high embodied carbon. 
While the added carbon from smaller technology nodes has increased performance per unit area, it may not increase energy efficiency by the required margins for many processors. 
Therefore, as shown in \autoref{tab:case-study}, an older server can become an attractive alternative to a new server, especially when its embodied carbon has been amortized in its initial expected lifespan. 
In the worst case, it will sort servers from the oldest and least efficient (serving base demand at all times) to the newest and most efficient (only used for infrequent peak demand).

While \SCI successfully incentivizes using older servers, it essentially provides an incentive to buy new servers, not use them, and use them only when they get old. 
We agree that the hardware should be used for longer and older servers should \emph{serve} a purpose, but serving base demand using them is not a sustainable strategy. 
Older hardware should be kept, but its high operational carbon should only be accepted during peak demand. 
Using \SCI to increase the operational carbon footprint is unnecessary and counter-intuitive. 
However, our analysis of metrics and case study findings requires that job scheduling decisions be decoupled from procurement decisions. 
When replacing an existing server, procurement teams can use \SCI to compare existing servers against available options and purchase only the servers with a lower \SCI. 
Note that the procurement for replacement differs from the procurement for new capabilities; if an emerging workload is critical but cannot be run on existing hardware, the purchase of new hardware will be \SCI-agnostic. 
However, once a set of servers has been procured, their embodied carbon has been emitted; the only goal should be to reduce the operational cost of running the servers.

\textbf{\TSCI} includes the datacenter-level embodied carbon and the operational carbon of the server running the job. 
This unified approach simplifies the allocation of carbon costs to users by aligning accounting and scheduling practices. However, embodied carbon estimates are highly uncertain due to variability in manufacturing processes, supply chain differences, and data quality issues. Relying on uncertain estimates for scheduling purposes risks making suboptimal decisions. Adding a noisy signal to an otherwise accurate accounting of operational carbon introduces errors that can cause incorrect resource allocation or prioritization decisions.
Furthermore, as discussed in \autoref{sec:illustration}, accurately computing and tracking \TSCI over time for large infrastructure, such as the public cloud, will incur significant overhead, prohibiting its use. 

\textbf{\OSCI} is the most effective metric for carbon-aware scheduling as operational carbon is the primary contributor in this scenario, and replacing existing hardware remains outside the scope at the timescales of scheduling decisions. Scheduling with \OSCI focuses exclusively on reducing operational emissions since this is the only component that can be directly optimized for the procured hardware. Hardware replacement decisions affecting embodied carbon are orthogonal to workload scheduling and should not influence this process.
Finally, using \OSCI reduces the operational costs for the infrastructure, on-premise or cloud, as it picks the most efficient hardware and does not succumb to the \emph{sunk carbon fallacy}. 

\vspace{-0.3cm}
\section*{Acknowledgements}
We thank the SoCC reviewers and our shepherd, Timothy Zhu, for their valuable comments, which improved the quality of this paper, and WattTime for access to carbon intensity data. 
This work was supported in part by the NSF grants CNS-2325956, CNS-2105494, and CNS-2213636, the NSF CAREER Award CCF-2326182, the NSERC grants RGPIN-2021-03714 and DGECR-202100462, a Sloan Research Fellowship, and an Intel Research Award. 

\balance

\end{document}